# Microstructure of Silicon Anodes in Solid-State Batteries - From Crystalline to Amorphous

Shamail Ahmed[1,2]*, Federico Rossi[3], Hanyu Huo[4,5], Johannes Haust[1,2], Franziska Hüppe[1,2], Jürgen Belz[1,2], Andreas Beyer[1,2], Jürgen Janek[3], and Kerstin Volz[1,2]*

[1] Department of Physics, Philipps-University Marburg, Hans Meerwein Str. 6, 35032 Marburg, Germany

[2] Marburg Center for Quantum Materials and Sustainable Technology (mar.quest) and Department of Physics, Philipps-University Marburg, 35043 Marburg, Germany

[3] Institute of Physical Chemistry, Justus Liebig University Giessen, Heinrich-Buff-Ring 17, D-35392 Giessen, Germany.

[4] Department of Materials Science and Engineering, University of Science and Technology of China, Hefei, Anhui 230026, China

[5] State Key Laboratory of Precision and Intelligent Chemistry, University of Science and Technology of China, Hefei, Anhui 230026, China

* Corresponding authors email: Shamail.ahmed@physik.uni-marburg.de; kerstin.volz@physik.uni-marburg.de

## Graphical abstract

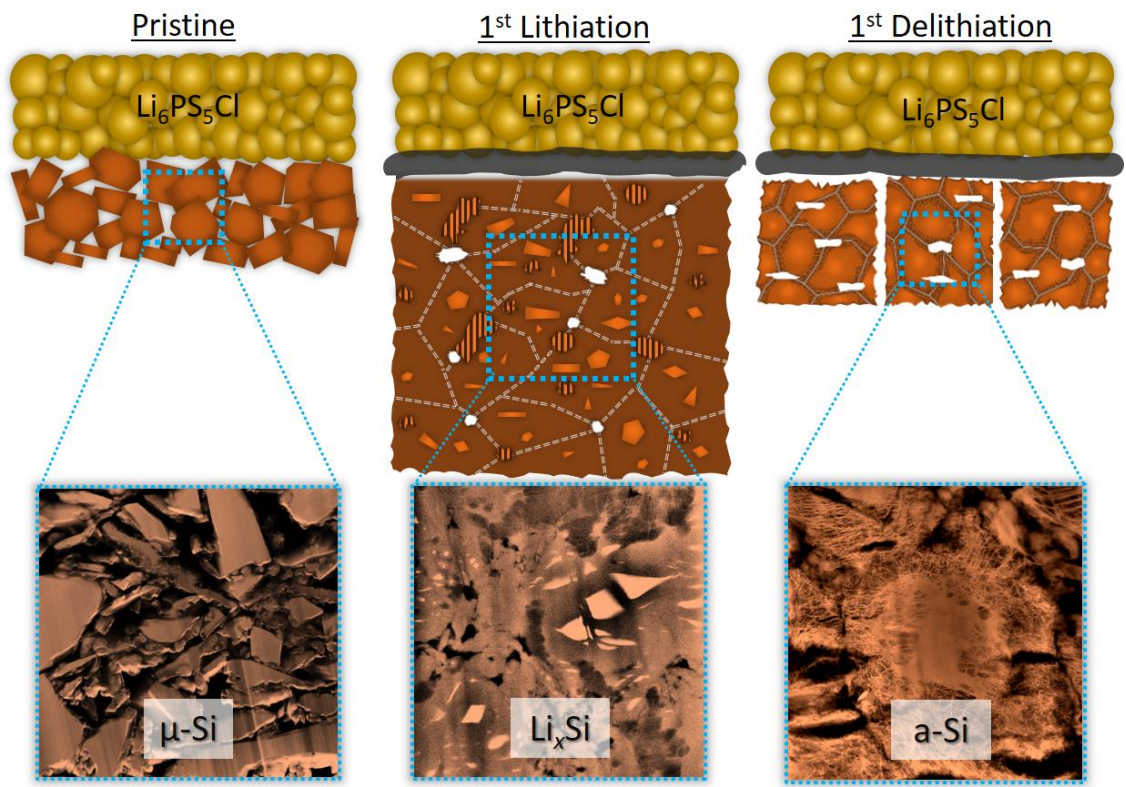

## Abstract

Silicon offers great promise as a potential anode active material and the optimum alternative to lithium metal in all-solid-state lithium-ion batteries. However, its practical application is limited by severe volume expansion (~300%) during lithiation, leading to cracking upon delithiation. In this study, we investigated the microstructural evolution of microcrystalline silicon electrodes in a solid-electrolyte-free environment using cryogenic scanning transmission electron microscopy (STEM) during electrochemical cycling. A controlled workflow prevents ambient exposure, and cryo-TEM ensures structural integrity. After the first lithiation, the electrode shows a heterogeneous mix of crystalline $Li_{15}Si_4$, various amorphous $Li_xSi$ phases, and residual crystalline silicon. After delithiation, the structure becomes predominantly amorphous with thread-like features and minimal remaining crystallinity. By the 10$^{th}$ delithiation, the microstructure is more uniform, with thread-like regions mainly at grain boundaries. Our results reveal that starting from a crystalline phase, a stationary microstructure emerges in bulk silicon only after several cycles. Thus, to have a more controlled behavior of the electrode and minimize cracking, the starting material should be carefully chosen along with an optimized electrode architecture to help stabilize the microstructure throughout cycling.

## Introduction

All-solid-state batteries (ASSBs) may show greater safety and higher energy density than lithium-ion batteries (LIBs), along with enhanced thermal stability and faster charging capabilities.[1] This will depend to a large extent on the successful utilization of a high-capacity anode. While the lithium metal anode is still plagued with stability issues, even in the solid-state environment, silicon is recently been considered a potential alternative anode material for enhancing the energy density of rechargeable ASSBs, due to its exceptionally high theoretical capacity ($q_{th}$ ($Li_{15}Si_4$) ≈ 3600 mAh g$^{−1}$; $q_{th}$ ($Li_{4.4}Si$) ≈ 4200 mAh g$^{−1}$) and low lithiation potential ($E$ < 0.35 V vs. Li$^+$/Li). Additionally, it is affordable, naturally abundant, and non-toxic.[2–6] It helps mitigate the formation of lithium dendrites, a significant challenge associated with lithium metal anodes.[7–9]

However, achieving a stable specific capacity is challenging owing to silicon´s high volumetric expansion ($\Delta V/V$ ≈ 300%) during lithiation and delithiation.[10,11] In liquid electrolyte-based lithium-ion batteries (LIBs), liquid electrolytes can penetrate the gaps between silicon particles and wet the surfaces of silicon particles completely. The expansion and contraction of silicon particles during charge-discharge cycles repeatedly lead to the cracking and local pulverization of silicon particles, and exposure of fresh surfaces, ultimately leading to continuous growth and thickening of the solid electrolyte interphase (SEI) layer that impedes the transport of lithium-ions.[3,12,13] Nevertheless, silicon is increasingly used in liquid-based commercial LIBs, albeit in limited quantities and typically in composite forms.[14]

When it comes to ASSBs, one of the commonly applied electrode configurations of the silicon anode is the silicon/solid electrolyte composite electrode (note as "3D-electrode" in the following), following the liquid electrolyte approach of dispersed electrode particles.[15–17] However, it has recently been found that the solid electrolyte interface (SEI) causes severe resistance build-up in silicon/$Li_6PS_5Cl$ composite anodes. A promising alternative to silicon/solid-electrolyte composite electrodes appears to be a thick and compact solid electrolyte-free (SE-free) silicon electrode layer (noted as "2D-interface electrode" in the following), which incorporates >99 wt% crystalline "micro-silicon" (μ-silicon) particles and a trace amount of binder, demonstrating improved capacity retention.[17–19] Although the aforementioned problems with liquid electrolyte-based batteries and ASSBs employing silicon/SE composite anodes occur less in the 2D-interface electrode configuration, their performance is still far from ideal, and several issues need to be understood and resolved before they can be employed practically in ASSBs. Two of the most prominent issues about silicon 2D-interface electrodes are silicon/SE-interface delamination and crack formation, particularly after delithiation.[17]

Usually, a severe cracking has been observed and reported after delithiation in the electrolyte-free silicon electrodes.[20–24] Nelson et al.[25] employed X-ray computed micro-tomography (XCT) to demonstrate the formation and propagation of mud-type channel cracks (cracks occurring between silicon domains) and the delamination of the silicon/SE interface across micro-to-macro scales during charge and discharge cycles. Both issues stem from morphological changes in the bulk structure of crystalline μ-silicon particles within the 2D-interface electrodes. However, the mechanisms of lithiation and delithiation, the influence of impurities, and the associated microstructural changes such as crystalline-to-amorphous transformations, the distribution of various lithium-containing phases, surface evolution, and the textural characteristics of lithiated and delithiated amorphous phase are not yet fully understood. Understanding these phenomena necessitates a detailed understanding of the nano-scale evolution of the bulk structure of crystalline μ-silicon particles during charge and discharge cycles. Furthermore, studying the morphological changes of μ-silicon during lithiation and delithiation in a SE-free environment could provide valuable insights for designing improved silicon/SE composite electrodes.

Scanning/transmission electron microscopy (S/TEM) is a powerful tool for uncovering nanostructures and has been effectively utilized to address critical aspects of battery materials.[26–32] Numerous *in situ*[33–40] and *ex situ* [35,41–43] S/TEM studies have been conducted on silicon anode materials; however, to the best of our knowledge, no S/TEM study has specifically explored the bulk structure of originally crystalline silicon in the ASSB environment, particularly in the 2D-interface electrode configuration. Conducting *post-mortem* S/TEM studies on silicon anodes is highly challenging, as such samples often require specialized target preparation using cryogenic focused-ion beam (FIB) techniques, inert gas sample transfers between several instruments, and cryogenic S/TEM analysis. On the other hand, it is difficult to replicate the ASSB environment in an *in-situ* experiment in an S/TEM.

Here we present and use a workflow to address the air, moisture, and ion and electron beam sensitivities of ASSB silicon anode samples. Silicon electrodes are retrieved in an argon-filled glovebox and transferred to the plasma FIB (PFIB) under an argon atmosphere. These samples are thinned and polished at -190 °C using Xe- and Ar-ion plasma beams in a PFIB and returned to the glove box. They are then mounted on a double-tilt inert-gas transfer holder for safe transfer to the S/TEM, where most measurements are conducted under cryogenic conditions. The silicon 2D-interface electrodes, containing crystalline μ-silicon particles, were cycled with a $Li_6PS_5Cl$ SE against an In/InLi alloy as a counter electrode. Following the outlined workflow, S/TEM samples of the silicon electrodes are prepared after the first lithiation, first delithiation, and tenth delithiation. These samples are analyzed using STEM high-angle annular dark field (HAADF) imaging, energy-dispersive X-ray spectroscopy (EDXS), electron energy loss spectroscopy (EELS), and four-dimensional scanning nano beam and precession electron diffraction (4D-SNBD and SPED), and critical aspects regarding morphological and microstructural changes in the bulk silicon are presented. Our results highlight that the crystalline-to-amorphous transformation during the first lithiation is complex, producing various lithium-containing silicon phases. A fully developed amorphous framework of originally μ-silicon particles forms only after multiple cycles. However, by that stage, the electrode's mechanical stability is already compromised, leading to cracks due to the uncontrolled crystalline-to-amorphous transformations that took place in the earlier cycles.

## Results and discussion

*Characterization of pristine μ-silicon particles*

In **Figure 1**, the morphology and composition of the pristine silicon are depicted. **Figure 1a** presents a secondary electron image, while **Figures 1b and 1c** display the corresponding EDXS maps illustrating the distribution of silicon and oxygen. This aligns with our previous study, where we demonstrated that the commercially available crystalline μ-silicon particles used in this work, also employed in our earlier research, typically have a thin $SiO_x$ coating on their surfaces.[17] **Figure 1d** shows a secondary electron image of the cross-section of a pellet pressed from crystalline μ-silicon particles, showing the large porosity in the microstructure. **Figure 1e** presents a STEM HAADF image of the pellet, also highlighting the multi-modal size distribution of the particles. The rough texture on the particle surfaces is attributed to the presence of a $SiO_x$ surface layer. The crystalline μ-silicon electrodes were cycled with $Li_6PS_5Cl$ solid electrolyte (SE) against an $In/(InLi)_x$ alloy as a counter electrode. Detailed information on typical cell configuration, cell fabrication, and comprehensive electrochemical characterization of silicon 2D-interface electrodes is provided in a separate study.[17] Details on the cell fabrication and sample extraction are provided in the methods section.

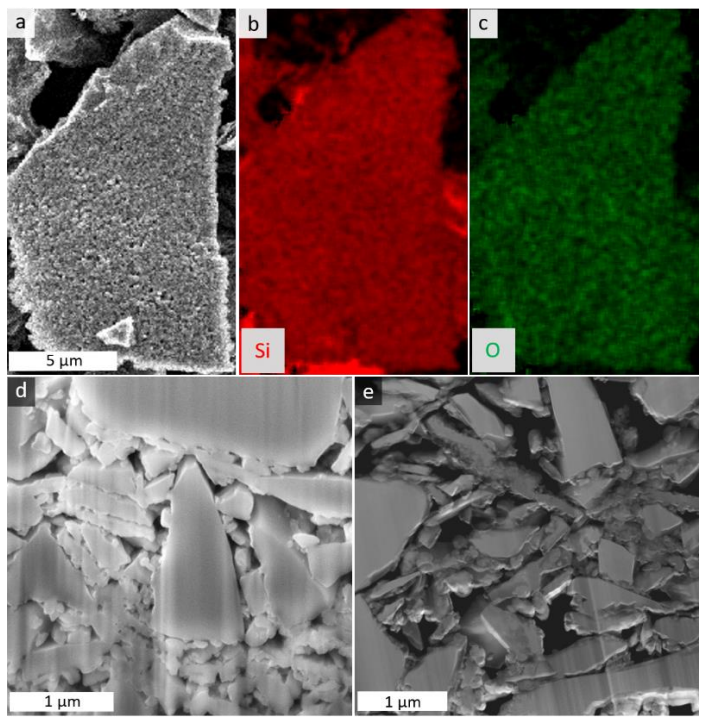

Figure 1: (a) Secondary electron image of a commercial crystalline μ-silicon particle. (b, c) EDXS maps show the distribution of silicon and oxygen. (d) Secondary electron image of the cross-section of a pressed crystalline μ-silicon particle pellet (e) STEM HAADF image of the pressed pellet.

*Characterization of Silicon after 1st lithiation*

**Figure S1** (provided in the supplementary information) presents the first lithiation curve along with the first and tenth delithiation curves. The cell was lithiated to a specific capacity of 3,314 mAh g$^{-1}$ (corresponding to Li$_x$Si with $x ≈ 3.47$), achieving a reversible capacity of 2537 mAh g$^{-1}$ ($x ≈ 2.66$) after the first delithiation. This indicates a significant loss of lithium in the silicon electrode following the first cycle. It is worth noting that the specific capacity and Li/Si ratio calculations are based on the measured mass of the silicon layer. Slight uncertainties in mass measurements may lead to slight errors. To prepare a TEM sample from the silicon electrode after the first lithiation, the cell was disassembled in an argon-filled glovebox, and the solid electrolyte (SE) layer was removed from the electrode. **Figure S2a** shows the electrode surface, where traces of Li$_6$PS$_5$Cl remained. **Figure S2b** provides a cryo-secondary electron image of the cross-section of the silicon electrode after the first lithiation. **Figure 2a** is a cryo-STEM HAADF image taken after the first lithiation. Compared to the pristine silicon electrode shown in **Figure 1e**, **Figures 2a and S2b** reveal a drastic reduction in porosity, with the original morphology of the µ-silicon particles completely transformed. The boundaries between µ-silicon particles have become relatively indistinct, and most particles appear to have fused.

**Figure 2b** presents a magnified cryo-STEM HAADF image of a different region, highlighting four distinct microstructural features observed after the first lithiation. The first feature, exemplarily enclosed by a green line and labeled as '1', displays a varying fraction of dendrite-like structures in the dark gray regions. The second feature consists of bright contrast areas, (exemplarily) outlined in red and labeled as '2'. The third feature, which constitutes the majority of the microstructure, comprises gray contrast regions marked with a dark blue dashed line and labeled as '3'. Finally, the fourth feature, (exemplarily) highlighted by an orange line and labeled as '4', corresponds to dark contrast areas representing pores. **Figure 2c** presents a magnified view of the area outlined by the black-dotted line in **Figure 2a**, highlighting dendrite-like features. **Figure S2c** displays a similar region with comparable characteristics. **Video SV1** (in the supplementary information) showcases cryo-STEM HAADF tomography along the horizontal axis of the dendrite-like region depicted in **Figure 2c**. These dendrite-like features represent a sheet-like microstructure, appearing sharp whenever their sheet axes align with the electron beam direction. The precise origin of these features, particularly their formation at specific locations, remains elusive at this point. This issue will be addressed in more detail later. **Figure 2d** is a virtual dark-field image of the area outlined by the dashed yellow line in **Figure 2a**, derived from the 4D SPED dataset acquired in this region. **Figures 2(i–iv)** present position-averaged SPED patterns corresponding to the highlighted regions in **Figure 2d**. The position-averaged SPED patterns 'i' and 'ii,' corresponding to the dendrite-like structures, reveal a mixture of crystalline and amorphous phases. The diffraction spots match well with the $Li_{15}Si_4$ structure, with several matched spots exemplarily marked in **Figures 2i and 2ii**. Pattern 'iii' corresponds to crystalline Si, while pattern 'iv' exhibits continuous rings in the diffraction pattern specific to an amorphous structure. **Figure 2e** displays a phase map showing the distribution of $Li_{15}Si_4$, Si, and amorphous phases in the microstructure. While the dendrite-like regions contain a mixture of amorphous and $Li_{15}Si_4$ phases, only the $Li_{15}Si_4$ phase is intentionally displayed in those regions of the phase map for clarity. Additionally, **Figure S2d** presents the radial pair distribution function (rPDF) derived from the amorphous diffraction patterns in **Figures 2ii and 2iv**. The similarity of these rPDFs suggests that the amorphous structure, overlapping with the crystalline $Li_{15}Si_4$ phase, in the region 'ii' of **Figure 2c** is comparable to the amorphous structure in region 'iv.'

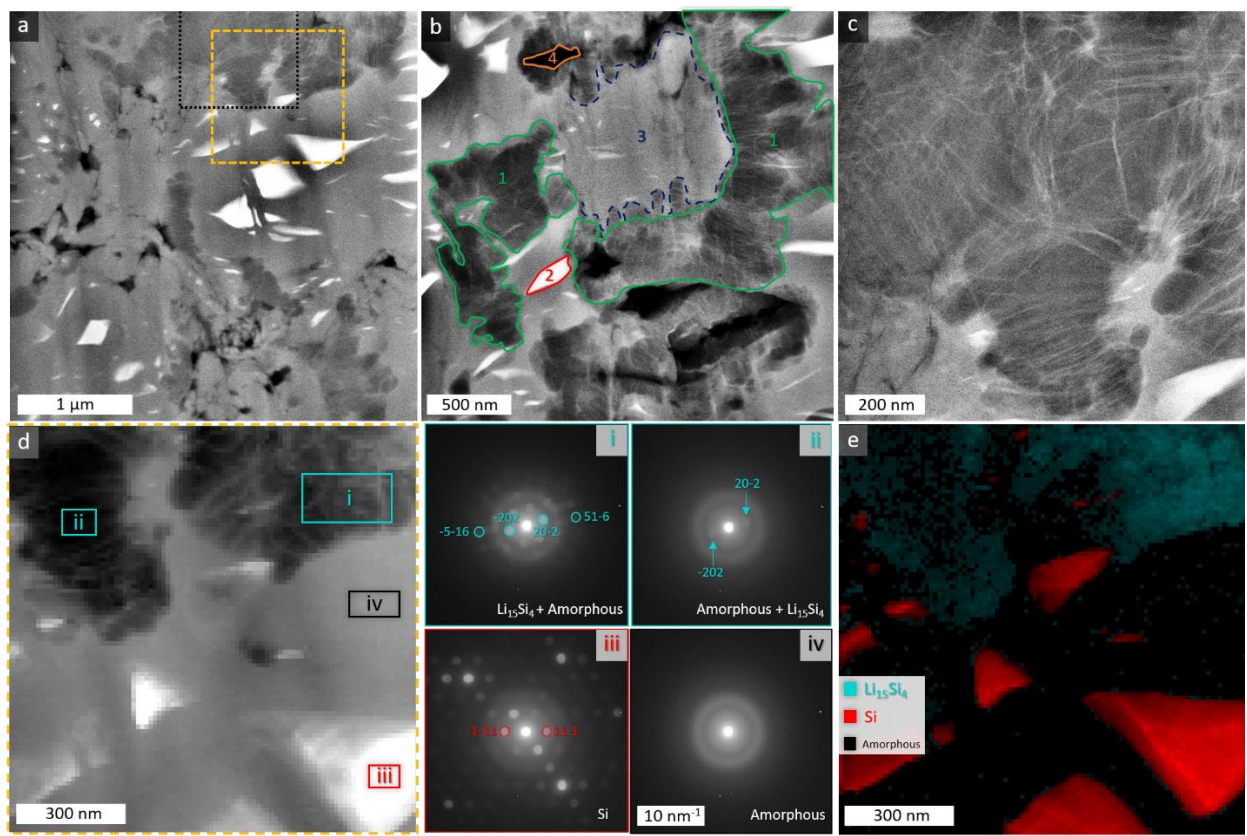

**Figure 2:** (a) Cryo-STEM HAADF image of the silicon electrode after the first lithiation. (b) Magnified cryo-STEM HAADF image exemplarily highlighting four distinct microstructural features: (1) dendrite-like structures (green line), (2) bright contrast areas (red line), (3) gray contrast regions (dark blue dashed line), and (4) pores (orange line). (c) is the magnified image of a region marked by an open-ended black-dotted rectangle in (a). Note that the magnified region in (c) extends beyond the boundaries of (a), focusing on additional details from the top portion of the open-ended black-dotted rectangle area depicted in (a). (d) Virtual dark-field image of the region outlined by the dashed yellow line in (a), derived from a 4D SPED dataset. (i–iv) Position-averaged SPED patterns corresponding to highlighted regions in (c). Some of the reflections respective to the structures identified are exemplarily marked in (i-iii). The rPDF plots corresponding to the diffraction patterns (ii) and (iv) are shown in Figure S2d. (e) Phase map showing the distribution of $Li_{15}Si_4$, Si, and amorphous phases in the microstructure.

A detailed analysis of simulated rPDFs for various $Li_xSi$ alloys reveals qualitatively similar rPDFs with broad peaks across different compositions[17] making it difficult to accurately determine the exact stoichiometry of the $Li_xSi$ alloys present in these regions based solely on rPDFs. Analysis of the 4D SPED datasets indicates that microstructural features 1, 2, and 3 exhibit significant differences, suggesting they contain varying amounts of lithium. Therefore, it is crucial to examine the chemical composition of these regions.

**Figure 3a** presents a cryo-STEM HAADF image, while **Figure 3b** displays a cryo-EDXS map illustrating the distribution of silicon and oxygen. A notable concentration of oxygen is observed near the pores and along the surfaces or grain boundaries of the particles. To analyze the Li/Si ratio, cryo-STEM valence electron EELS mapping is employed. Danet et al.[44] identified the plasmon energy values for various lithiated silicon compounds, which are used in this study to derive a fit for the plasmon energies corresponding to lithium content in Li$_x$Si alloys, ranging from $x$ = 0 to $x$ = 4.4. **Figure S3** presents the fitting curve and the associated equation for determining the plasmon energies of various Li$_x$Si alloys. It is important to recognize the practical challenges of using core-loss lithium-K and silicon-L$_{2,3}$ edges to determine Li/Si ratios, as highlighted by Danet et al.[44] Variations in sample thickness introduce multiple plasmon peaks (first, second, third, etc.), complicating background fitting.[44] In thicker sample regions, the fourth plasmon peaks from Li$_x$Si can significantly interfere with the lithium K-edge, as these plasmon peaks are located near the lithium K-edge. Due to the complex microstructure, with features of varying densities (as shown in **Figures 2(a & b), S2b, and 3a**), achieving uniform sample thickness across the TEM lamella is highly challenging. Large thickness variations, combined with the moderately low-dose electron beam used, make it difficult to proportionally excite the lithium-K and silicon-L$_{2,3}$ edges simultaneously. Given these constraints, valence electron EELS provides a more suitable method for analyzing Li/Si ratios in the silicon electrodes with their complex microstructure. **Figure 3c** presents a cryo-STEM HAADF image of a selected region, while **Figure 3d** shows the corresponding cryo-valence electron EELS map, highlighting the Li/Si ratios. We fitted Lorentzian functions using non-linear least squares (NLLS) fitting to find out the positions of the plasmon peaks. The microstructural features identified in **Figure 2b** are also marked for reference in **Figures 3(c & d)**. The dendrite-like regions (e.g., region '1') exhibit the highest lithium content, although the lithium concentration varies significantly in these regions. This variability aligns with observations from **Figure 2c** and the SPED patterns in **Figures 2i and 2ii**, which indicate a mixture of crystalline Li$_{15}$Si$_4$ and amorphous Li$_x$Si phases in projection. Typical Li/Si ratios in these regions range from 2.8 to 4.3. Closer examination of **Figures 2(a & b) and 3(a & c)** reveals that these high-lithium containing, dendrite-like regions are predominantly (but no exclusively) located somehow near the grain boundary regions gradually extending inwards of the particles. Regions marked as '2' contain no detectable lithium. As shown in **Figures 2c, 2iii, and 3d**, these regions correspond to pure crystalline silicon that remained un-lithiated, often located in the particle cores. Regions marked as '3' display an average Li/Si ratio of approximately 1.8, indicating moderate lithiation. The average lithium content per silicon in **Figure 3d** (including pores) is approximately 2.2, significantly lower than the expected 3.47 lithium per silicon value achieved during lithiation and the observed reversible capacity corresponding to a Li/Si ratio of about 2.66. A closer inspection of **Figure S2b** reveals that a substantial portion of the electrode remains un-lithiated, consisting of pure crystalline silicon along with regions similar to those labeled as '3' in **Figures 2b, 3c, and 3d**, which exhibit only moderate lithiation. The presence of pure crystalline silicon indicates that the relatively low amount of lithium (less than expected) in the microstructure is real and cannot be attributed to artifacts from sample preparation, highlighting that there are regions where lithium ions were not able to penetrate silicon during lithiation.

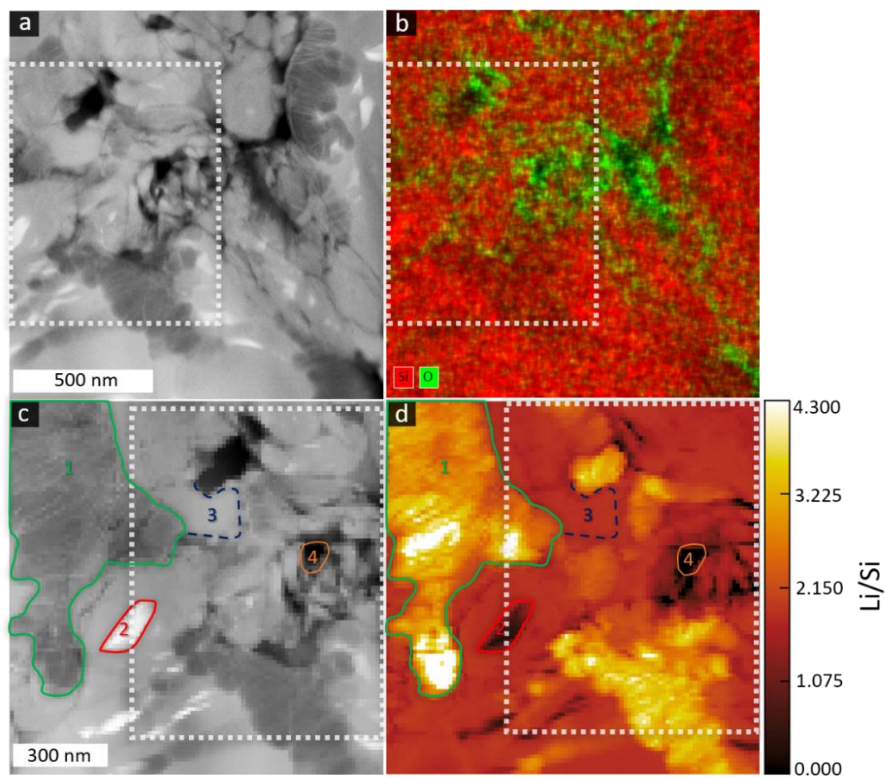

**Figure 3:** (a) Cryo-STEM HAADF image of the silicon electrode after the first lithiation. (b) Cryo-EDXS map showing the distribution of silicon and oxygen, with a notable concentration of oxygen near pores and particle surfaces. (c) Cryo-STEM HAADF image of a selected region, exemplarily marking distinct microstructural features as shown in Figure 2b. (d) Cryo valence electron EELS map of the same region, displaying the Li/Si ratios. The white dashed-dotted rectangle in (a), (b), (c), and (d) highlights the shared region across these images.

Several factors could explain the observed under-lithiation of μ-silicon particles. First, a considerable amount of lithium is consumed due to SEI formation at the 2D interface present between the silicon electrode and $Li_6PS_5Cl$. Second, as shown in **Figures 1(a–c)** and previously reported by Huo *et al.*[17] a significant oxygen content exists on the surfaces of the crystalline μ-silicon particles. To qualitatively assess lithium loss at grain boundaries in the form of $Li_2O$ and $Li_ySiO_x$, valence electron EELS was employed (shown in **Figure S4**). The plasmon peaks for $Li_2O$ and $SiO_2$ are located at 19 eV and 22 eV,[41,45] respectively. While this range could theoretically be used to map these compounds and their mixtures, the dominant presence of $Li_xSi$ in the microstructure and potential overlap from projected signals of $Li_xSi$, $Li_2O$, and $Li_ySiO_x$ make mapping within the 19–22 eV range challenging due to plasmon peak interference form $Li_xSi$ compounds (which lie between 12.74 eV and 16.83 eV).[44]

To address this issue, the second plasmon range for these compounds was utilized instead. The spectrum at each pixel position was subjected to Lucy-Richardson deconvolution, and a cryo-valence electron EELS map was constructed using a (42 ± 2) eV integration window. This map, shown in **Figure S4b**, highlights significant $Li_2O$ and lithium-silicate compound concentrations at grain boundaries. EDXS mapping from the shared region between **Figures S4b and 3b** also confirms the presence of oxygen at pores and some grain boundaries. An overpotential observed during 1$^{st}$ lithiation in **Figure S1** may indirectly indicate the formation of such compounds. Additionally, the map in **Figure S4b** reveals hidden grain boundaries, which are often obscured by particle expansion. It is worth noting that such a map may not be reliably obtained using the lithium-K edge, particularly in silicon after first lithiation, as lithium is present in all candidate compounds ($Li_xSi$, $Li_2O$, and $Li_ySiO_x$) and exhibits only minor lithium-K edge structural differences in these phases. Additionally, the potential overlap of these compounds, particularly at the grain boundaries due to projection effects, complicates the deconvolution and differentiation of various individual lithium-K edge spectra.

The third factor contributing to the discrepancy between the expected and measured lithium content in the structure is that the fitting of plasmon energies for different $Li_xSi$ compounds is based on a limited set of $Li_xSi$ compositions. Expanding the dataset of plasmon energies for $Li_xSi$ compounds, as shown by Danet et al.,[44] would enhance the reliability of Li/Si ratio quantification. Furthermore, as the lithium content increases, the incremental decrease in electron density of the $Li_xSi$ compounds diminishes, causing the relative flattening of the curve as it approaches $Li_{4.4}Si$ (as shown in **Figure S3**). Due to this flattening effect, the mapping sensitivity decreases in regions with high lithium content, making it difficult to differentiate areas with subtle variations in high lithium concentrations.

Lastly, the presence of a thin passivation layer of $Li_2O$ on the TEM lamella may contribute to the observed discrepancy. Despite handling the samples in a glovebox and transporting them between microscopes under an inert gas atmosphere, the formation of such layers, particularly in the case of highly lithiated Si or lithium metal, is likely unavoidable.[46]

Considering the practical constraints and the absence (to the best of our knowledge) of prior studies quantifying Li/Si ratios in a SE-free ASSB environment using STEM, we regard these findings as crucial for advancing our understanding of the microstructure of lithiated µ-silicon particles.

The results from silicon after the first lithiation indicate that the lithiation of µ-silicon in the 2D-interface-electrode configuration is inhomogeneous and incomplete, despite galvanostatically charging the electrode to the Li/Si ratio of approximately 3.47. The crystalline-to-amorphous transformation was also incomplete, with a substantial portion of silicon remaining as pure crystalline silicon. The dendrite-like structures with high lithium concentration are predominantly, though not exclusively, located near grain boundaries and comprise a variable mixture of $Li_{15}Si_4$ and amorphous phases. Their presence near grain boundaries or microparticle surfaces may be attributed to the lithiation process, which typically initiates at the particle surface. However, an examination of **Figures 2a** and **3a** reveals that these features are not uniformly distributed near all pores, grain boundaries, or particle surfaces. This suggests a more complex interplay involving the availability of lithium from various migration fronts of neighboring particles during lithiation. Additionally, specific silicon crystal facets may influence the formation of these structures. The observed sheet-like texture of the dendrite-like features may indicate that lithiation in these regions occurred preferentially along particular silicon planes. A more comprehensive TEM investigation, involving pure microparticles exposed to varying degrees of lithiation along carefully chosen crystallographic directions, supported by theoretical calculations, may be required to fully elucidate this phenomenon. One such controlled study by Astrova et al.[47] revealed that the lithiation rate is highest along the [110] planes in silicon. However, this study did not include a detailed microstructural analysis using TEM.

The grain boundaries also contain a significant amount of $Li_2O$, which may play a key role in the under-lithiation of the μ-silicon particles. Sivonxay et al.[48] suggested that the presence of $SiO_2$ on the surface can impede lithium transport and act as an irreversible trap for lithium. While lithiation of $SiO_2$ may result in the partially reversible formation of $Li_xSiO_y$ compounds,[49] we believe that the presence of oxygen on the surface of μ-silicon particles has a significant role in hindering their full lithiation.

*Characterization of Silicon after the 1$^{st}$ delithiation*

**Figure S5a** displays the top surface of the silicon electrode after the first delithiation, revealing the presence of mud-type channel cracks. Nelson et al.[25] conducted an in-depth study detailing the dynamics of these cracks. **Figures S5(b & c)** present cross-sectional views of the silicon electrode during the lift-out process, while **Figure S5d** shows a magnified region of the cross-section. The magnified image highlights the partial recovery of porosity; however, the microstructure remains significantly altered compared to the pristine sample depicted in **Figure 1c**.

**Figure 4a** presents a cryo-STEM HAADF image of the silicon electrode after the first delithiation, showing structural differences compared to silicon after the first lithiation (**Figure 2a**). The delithiated sample exhibits porosity with distinct particle boundaries. **Figure 4b** provides a magnified cryo-STEM HAADF image of another region, revealing another distinct type of dendrite-like structure compared to the first lithiated sample (**Figure 2b**). The texture of the microstructure can be described as an irregular mesh of intertwined threads, reflecting its intricate and interconnected morphology. The particle core, enclosed by a dashed dark blue line, appears comparatively denser, while the areas near grain boundaries typically display a thread-like irregular mesh. **Figure 4c** is a virtual dark-field image constructed from a 4D scanning nano-beam diffraction (SNBD) dataset acquired in the region outlined by the dashed yellow line in **Figure 4a**. **Figures 4i and 4ii** show position-averaged SNBD patterns from regions marked as "i" and "ii" in **Figure 4c**. Both denser and less dense regions exhibit continuous rings in the diffraction patterns. **Figure S6a** provides the rPDFs for these two regions, both of which show characteristic peaks of the amorphous silicon structure.[50] Despite the structural differences observed in the ADF image, the variations in the PDFs remain subtle. It is important to note that two phases of the same disordered material may exhibit short-range order peaks at similar positions in the rPDFs, while their medium-range order peaks can still differ. These variations arise from their distinct pair-angle distribution functions (PADF), leading to differences in densities.[51,52] This aspect will be explored in a future study. Crystalline silicon is rarely found compared to silicon after first lithiation. An example of such a region is illustrated in **Figures S6b and S6c**.

**Videos SV2 and SV3** present cryo-STEM HAADF tomographs of one of the thread-like irregularly meshed regions depicted in **Figures S7a and S7b**. The videos showcase the region tilted about the horizontal and vertical axes, respectively. Interestingly, these thread-like features also appear diffuse when the sample is tilted, suggesting that they are sheet-like structures. When the sheet axis aligns with the electron beam, these features appear sharp and well-defined. Within these sheet-like structures, a porous morphology is visible, as shown in **Figure S7b**. These sheet-like structures support our earlier speculation that lithiation may preferentially occur along specific crystallographic planes and directions in certain regions.

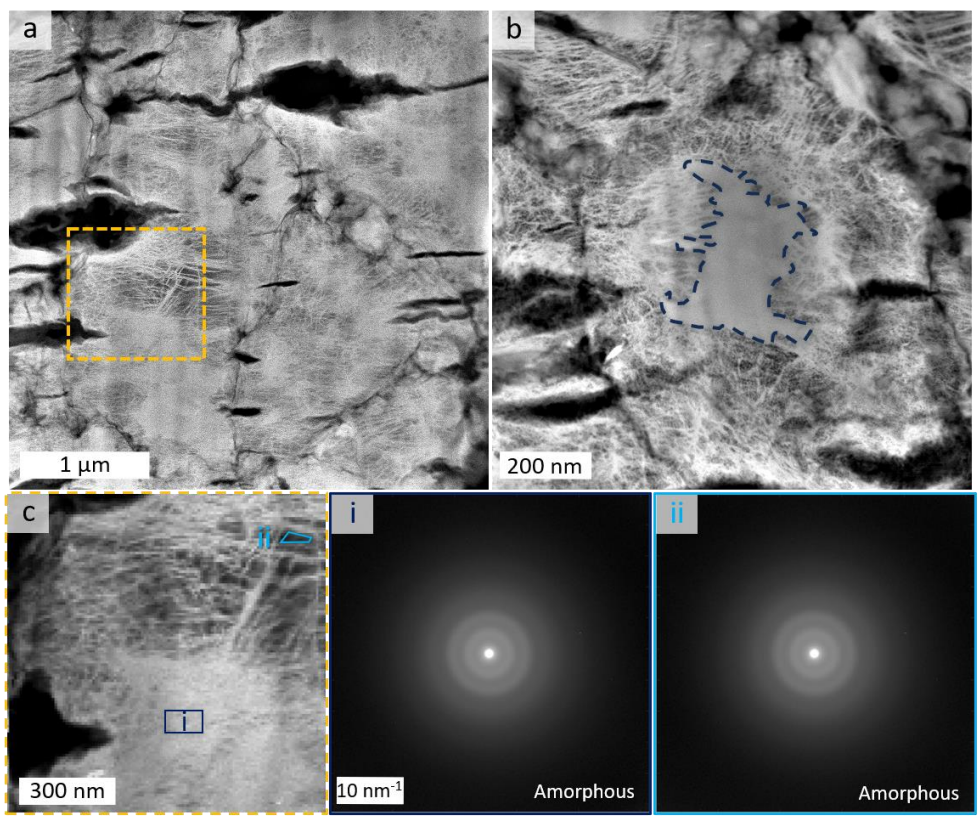

**Figure 4:** (a) Cryo-STEM HAADF image of the silicon electrode after the first delithiation. (b) Magnified cryo-STEM HAADF image of another region. (c) Virtual dark-field image constructed from a 4D SNBD dataset of the region outlined by the dashed yellow line in (a). (i and ii) Position-averaged SNBD patterns from the denser 'i' and less dense 'ii' silicon amorphous regions marked in (c), respectively. The rPDF plots corresponding to the diffraction patterns (ii) and (ii) are shown in Figure S6a.

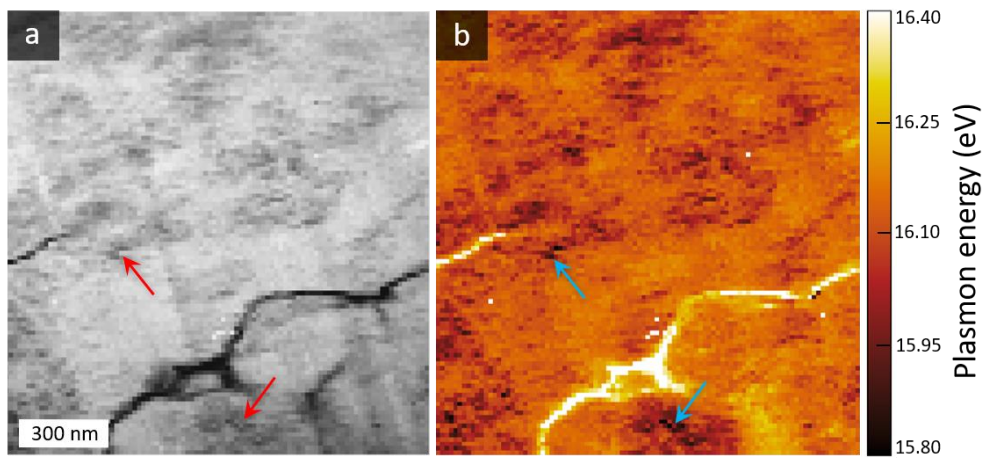

**Figure 5:** (a) Cryo-STEM HAADF image of a region after the first delithiation. (b) Cryo-STEM valence EELS map showing the plasmon peak positions across the area. The thread-like regions are exemplarily marked by red and blue arrows in (a and b).

**Figure 5a** shows a cryo-STEM HAADF image of a region after the first lithiation, while **Figure 5b** provides a cryo-STEM valence EELS map indicating the positions of the first plasmon peaks across the area. Most of the structure exhibits plasmon peak positions typical of amorphous silicon (16.1

eV - 16.7 eV).[53,54] However, in the thread-like regions (indicated by red arrows), the plasmon peak positions are typically below 16 eV. Although no significant lithium-K edge signals were detected in these areas, it remains uncertain whether the lower plasmon values are due to trace amounts of residual (and undetectable) lithium or the lower density of the structure in these regions. The average plasmon peak position, excluding regions between grain boundaries, was found to be at 16.12 eV, with the lowest plasmon peak position recorded at 15.80 eV. **Figure S8a** presents a cryo-STEM HAADF image of a region after the first delithiation. As no detectable lithium is found within the bulk silicon structure after the first delithiation the lithium-K edge can also be utilized to map lithium-containing compounds at grain boundaries. **Figure S8b** shows the corresponding cryo-STEM EELS map, constructed by integrating the energy window around the lithium-K edge (60 eV ± 2 eV) after zero-loss peak deconvolution and subsequent background subtraction. **Figure S8c** is a cryo-STEM valence EELS map generated using the range corresponding to the second plasmons of the $Li_2O$-$SiO_2$ mixtures (42 eV ± 2 eV). The similarity between **Figures S8b and S8c** confirms the results from the first lithiated sample (**Figure S4b**) demonstrating that plasmon mapping can effectively trace $Li_2O$ and lithium-silicate compounds, also highlighting that lithium is present where oxygen is generally present in these samples.

After the first cycle, the structure is predominantly amorphous, with two distinct regions identified: a denser amorphous structure generally (but not exclusively) more dominant at the particle cores, and irregular thread-like meshed regions observed generally (but not exclusively) near the grain boundaries. While the rPDFs of both microstructures qualitatively match the typical characteristics of amorphous silicon, subtle differences are evident in the peak ratios and peak positions. Additionally, the plasmon peak positions of these two structures differ slightly. Following delithiation, the structure contracts, recovering some porosity. This porosity makes it easier to trace trapped lithium, likely present as $Li_2O$ and $Li_ySiO_x$, compared to the first lithiated sample, where these compounds are also present but are compressed between the lithiated silicon particles.

*Characterization of Silicon after the 10$^{th}$ delithiation*

**Figure S9a** shows an SEM image of the surface of the silicon electrode after 10 cycles, while **Figure S9b** provides an SEM image of the cross-section during the lift-out process. Mud-type channel cracks are again evident after 10 cycles. **Figure 6a** displays a cryo-STEM HAADF image of the electrode after 10 cycles, with **Figure 6b** showing a magnified view of the region enclosed by the dashed yellow line in **Figure 6a.** The irregular thread-like meshed regions are now predominantly visible at the grain boundaries. **Figure S10a** is a virtual dark-field image derived from a 4D SPED dataset, with **Figure S10b** showing a position-averaged SPED pattern from the region marked in **Figure S10a**. **Figure S10c** provides the corresponding rPDF, which matches that of a typical amorphous silicon structure.[50] No traces of crystalline silicon were detected. **Figure 6c** is a cryo-valence EELS map of the region shown in **Figure 6b**, displaying the plasmon peak's position at each pixel. The map reveals an average plasmon peak position of 16.62 eV, with a minimum peak position at 16.5 eV, excluding grain boundaries. These values are notably higher than those obtained from the first delithiated sample. **Figures S10d and S10e** were generated using the lithium-K edge and the range corresponding to the second plasmon peaks of the $Li_2O$-$SiO_2$ mixture (42 eV ± 2 eV), respectively, confirming the presence of $Li_2O$ and lithium-silicate compounds at the grain boundaries.

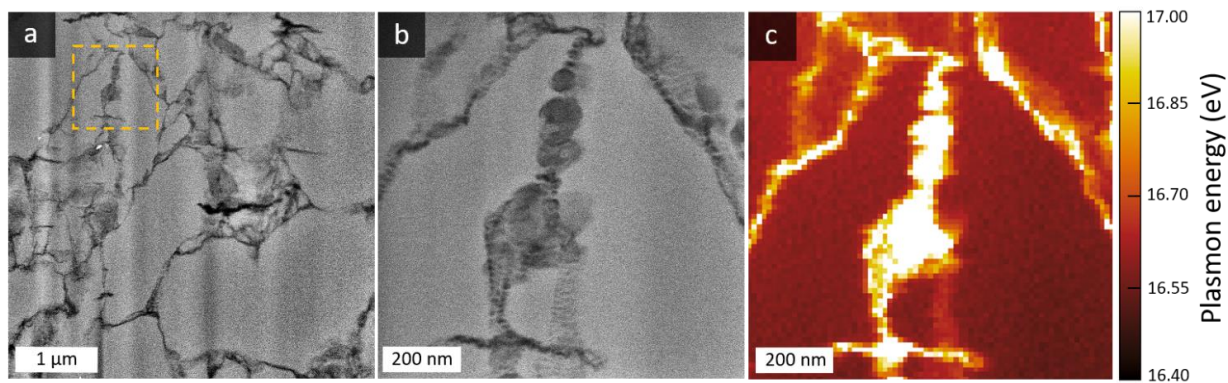

**Figure 6:** (a) Cryo-STEM HAADF image of the silicon electrode after 10 cycles. (b) Magnified view of the region enclosed by the dashed yellow line in (a). (c) Cryo-valence EELS map of the same region, illustrating the first plasmon peak positions.

## Conclusions

The specialized sample handling, transfer, preparation, and measurement procedures developed and adopted allowed for characterizing silicon electrodes extracted from ASSB cells with STEM. The silicon electrodes in this study consisted originally of crystalline μ-silicon particles with a $SiO_x$ surface layer, exhibiting rough textures, porosity, and a multi-modal size distribution. The first lithiation caused an inhomogeneous and incomplete lithium distribution, leaving a notable crystalline fraction. Regions with a dendrite-like microstructure and high lithium concentration near grain boundaries contained a mixture of $Li_{15}Si_4$ and amorphous $Li_xSi$ phases, while $Li_2O$ and lithium-silicates at grain boundaries appear to have hindered complete lithiation. The $SiO_x$ layer on μ-silicon surfaces is believed to impede lithium transport and trap lithium. After the first cycle, silicon becomes predominantly amorphous, with denser regions at particle cores and thread-like features at grain boundaries, differentiated by subtle variations in rPDFs and plasmon peak positions. Lithium is primarily localized at grain boundaries, likely as $Li_2O$ and $Li_ySiO_x$. Although a significant crystalline silicon fraction is still found after the first lithiation, it transitions to an amorphous state after delithiation, with crystalline silicon becoming rarely found. The bulk silicon structure demonstrates reversible lithiation and delithiation, with no trapped lithium detected except at oxygen-rich grain boundaries. After 10 cycles, the electrode shows amorphous silicon grains and thread-like regions at grain boundaries, with the bulk grain structure becoming more uniform. Lithium compounds, including $Li_2O$ and lithium-silicates, remain concentrated at the grain boundaries. Overall, this study reveals critical aspects of silicon lithiation behavior in ASSBs, identifying the key phases formed (e.g., $Li_{15}Si_4$, $Li_xSi$, $Li_2O$), their spatial distribution (such as at grain boundaries or within bulk grains), and the structural evolution across lithiation and delithiation.

Regarding the electrochemical implications of these findings, our study reveals that a stable amorphous electrode architecture in silicon-based electrodes develops only after multiple electrochemical cycles. Attempting to engineer a crystalline structure for such electrodes is likely not viable, as the initial crystalline-to-amorphous transformation occurs rather anisotropically, disrupting the electrode architecture. Instead, a formation cycle is required to generate a stable amorphous phase, which can then be extracted and incorporated into redesigned electrodes. However, the practicality and scalability of this approach remain uncertain. Recently, Lin et al.[55] attempted to engineer silicon electrodes using silicon pre-amorphized via delithiation and reported a significant reduction in electrode cracking. Starting directly with an amorphous material seems a more viable alternative. While this 'does not eliminate silicon's inherent volume fluctuations' during lithiation and delithiation, it mitigates the initial anisotropic and destructive phase transition from crystalline to amorphous states that could compromise structural integrity. It is speculated that if amorphous silicon is used as the starting material, the resulting amorphous phase, while still undergoing expansion during lithiation, would exhibit more uniform and predictable mechanical behavior. This allows for targeted design strategies, such as incorporating buffer spaces or flexible composites/layers, to better accommodate volume changes and enhance electrode durability without sacrificing electrochemical performance, as recently demonstrated by Huang et al.[56]

Additionally, while it is known that pre-lithiated crystalline μ-silicon particles perform better than their untreated counterparts,[57] we speculate that the performance may further be improved when pre-lithiated amorphous silicon particles are used. This speculation is based on our analysis of the samples after the first lithiation that the incomplete amorphization of the crystalline microparticles during the initial lithiation step leaves residual crystallinity and makes the material susceptible to further anisotropic local crystalline to amorphous phase transformation. Although crystalline $Li_{3.75}Si$ may still form locally under deep lithiation later during battery operation, we think that suppressing the first-cycle anisotropic crystalline-to-amorphous transformation at the bulk level significantly improves the overall mechanical and electrochemical stability of the electrode.

## Methods

*Materials and electrode preparation*

LPSCl particles (Posco JK Solid Solution) with a grain size of about 5 μm were used as received for the separator. Silicon particles (μ-silicon, 1–5 μm, 99.9% purity, Alfa Aesar) were dried in a Büchi furnace at 80 °C overnight before use. Polyvinylidene fluoride binder (PVDF) (Kynar HSV-900) was used as received as the binder for the SE-free silicon sheet. Indium foil (99.99%, 100 μm, Alfa Aesar) and lithium foil (99.9%, 100 μm, Albemarle) were used as received for the In/InLi alloy anodes.

SE-free silicon sheet anodes were prepared by casting a slurry (solid ratio of 56 wt%) of 99.5 wt% silicon particles, 0.5 wt% PVDF, and N-methyl-2-pyrrolidone (NMP) onto a copper collector using a doctor blade (30 μm). The sheets were vacuum-dried at 80 °C overnight, and 9.6 mm electrode discs were punched. The silicon loading was 1.8 mg cm$^{-2}$, with a thickness of 11.7 μm, and a porosity of about 33%.

*Cell fabrication and electrochemical characterization*

To assemble the cells, 80 mg of 5 μm LPSCl particles were densified by hand-pressing, followed by uniaxial pressing at 380 MPa after placing the SE-free silicon sheet disc on one side of the separator. An indium foil (∅ = 9 mm) and a lithium foil (∅ = 8 mm) were added on the other side of the separator to form the In/InLi alloy as the counter electrode.

The cells were fixed within a stainless-steel frame to maintain a constant pressure of 50 MPa. Long-term charge/discharge testing was conducted using a MACCOR battery cycler. Galvanostatic cycling was performed between −0.6 V and 1.0 V at a current of 500 μA (C/10 rate), with a 5-minute rest period after each charge and discharge step. Over 20 cells were tested, all showing consistent performance.

*Scanning electron microscopy (SEM) and sample preparation*

Due to the air and moisture sensitivity of ASSB samples, the cells were disassembled in an argon-filled glovebox, where the silicon electrodes were carefully retrieved. The ~800 μm thick LPSCl layer is removed from the silicon electrode before lamella preparation to prevent self-discharge during transport of samples between different facilities and sample storage. Additionally, detaching the thick LPSCl layer helps mitigate sample charging, particularly at cryogenic temperatures during preparation, and reduces the risk of silicon electrode interaction with LPSCl and localized heating at the silicon/LPSCl interface, which could potentially affect the lithiated silicon electrode. Special care is taken to minimize the time between LPSCl layer removal and TEM lamella preparation for the first lithiated sample. Due to the formation of mud-type channel cracks and slight delamination of the LPSCl layer from the silicon electrode, removing this layer from delithiated electrodes is relatively straightforward. However, detaching the LPSCl layer from the first lithiated silicon electrode is significantly more challenging. In this case, the removal process induces cracks in the electrode that are otherwise absent in cross-sectional views when the LPSCl layer remains intact.[17] Most of the LPSCl-peeling-induced cracks remain hidden and only become visible after a several-micrometer-thick bulk electrode chunk, which is attached to the TEM grid, is milled down further. These cracks drastically reduce the likelihood of successfully preparing a TEM lamella from the first lithiated sample.

The samples, after removing the LPSCl layer, were transferred under an argon atmosphere to a plasma FIB (PFIB) chamber using a shuttle. Utilizing the Helios 5 Hydra CX PFIB, the specimens were thinned and polished at approximately -190 °C with Xe- and Ar-ion beams. The pristine silicon samples were prepared entirely at room temperature. SEM images were taken using the same SEM/PFIB. For every sample, initially, a thin tungsten layer was deposited using an electron beam, followed by the deposition of a 3–4 µm thick tungsten layer with a 12 kV Xe-ion beam at room temperature. For the first delithiated sample, the initial trenching and lift-out process was performed at room temperature, with subsequent thinning and polishing carried out at approximately -190 °C. In contrast, for the first lithiated samples, the initial trenching was also conducted at approximately -190 °C, while the lift-out process was performed at room temperature, followed by final thinning and polishing at approximately -190 °C. During the thinning process, 30 kV Xe- and Ar-ion beams were employed until the sample reached a thickness of about 300 nm, with final thinning performed using 5 kV ion beams until the sample reached a thickness of about 150-100 nm. Once prepared, the samples were returned under argon atmosphere to the argon-filled glovebox, where they were mounted onto a double-tilt Atmos Defend Holder from Melbuild. This specialized holder facilitates the secure transfer of the sample from the glovebox to the S/TEM while maintaining an argon atmosphere and achieving measurement temperatures of approximately -170 °C during measurements. The SEM EDXS were done using ThermoFischer Scientific's UltraDry silicon drift X-ray detector, and the data were analyzed using ThermoFischer Scientific's Pathfinder (version 2.8) software.

It is important to note that while the inert gas transfer and sample preparation methods used in this study are among the best available solutions, they are not entirely effective in preventing oxidation. Research on lithium metal suggests that even trace amounts of oxygen and moisture in ultra-high vacuum chambers can lead to reactions with lithium.[46,58] However, cryogenic sample preparation and measurements help slow down the kinetics of $Li_2O$ formation.[46] Although pure lithium metal was not examined in this study, it can be inferred that lithiated silicon may also undergo oxidation, albeit to a significantly lesser extent than lithium metal. In contrast, the oxidation of delithiated silicon samples is expected to be even more suppressed due to their lower lithium content with the adopted procedure.

*Scanning transmission electron microscopy (STEM)*

We employed cryo aberration-corrected scanning transmission electron microscopy (AC-STEM) for sample characterization, except for the pristine samples, which were analyzed at room temperature. In this paper, the term "cryo" preceding a measurement indicates that the measurement was conducted under cryogenic conditions. STEM HAADF imaging, EDXS, and EELS measurements were conducted at approximately -165 °C using a double-tilt Atmos Defend Holder from Melbuild in a double aberration-corrected JEOL 2200FS STEM operating at 200 kV. STEM EDXS measurements were performed using a Bruker XFlash 5060 silicon drift detector (SDD), and the data were analyzed with QUANTAX ESPRIT software (version 2.3). EELS datasets were recorded with a TVIPS TemCam-XF416 camera, and data analysis was performed using HyperSpy[59] and custom-developed scripts written in Python and MATLAB. Non-linear least squares (NLLS) fitting was executed with Gatan's Digital Micrograph (version 3.40.2804.0) software. For mapping $Li_2O$ and lithium-silicate compounds at grain boundaries using second plasmons and lithium-K edges, the zero-loss peak was deconvoluted from the spectra using Lucy-Richardson deconvolution. STEM tomography was performed at approximately -165 °C using a double-tilt Atmos Defend Holder from Melbuild, with a tilt range of -13° to 13° in 1° increments. The images for the tomography are aligned using the 'Linear stack with alignment SIFT' plugin available in ImageJ (version 1.54k) software.

Scanning nano-beam diffraction (SNBD) and scanning precession electron diffraction (SPED) datasets were acquired at room temperature using a JEOL 3010 TEM operating at 300 kV and equipped with a NanoMegas P2010 system, utilizing a double-tilt Atmos Defend Holder from Melbuild. To minimize the electron dose, the smallest spot size and condenser lens aperture were employed. With the $LaB_6$ electron gun of the JEOL 3010 together with the parameters used, a beam diameter of approximately 4 nm was achieved without electron beam precession, while a diameter of around 8 nm was obtained with 0.4° electron beam precession. SNBD and SPED datasets were recorded with a TVIPS TemCam-XF416 camera initially as video streams. These initial video streams are converted to rectangular 4D datasets and then to .bloc files using custom-written scripts in Python. The .bloc files were analyzed with ASTAR (version 2) software packages from NanoMegas for crystal structure identification and phase mapping. The radial pair distribution function (rPDF) analysis was done using custom-written codes in Python and MATLAB.

*TEM sample statistics*

A total of seven TEM samples were prepared from several different lithiated silicon electrode samples. However, due to the challenges associated with lithiated samples (as previously mentioned), only one sample reached the optimal quality required for practical TEM analysis.

Analyzed regions covered approximately 10 μm × 10 μm sample area. In the lithiated samples, individual μ-silicon particles are difficult to distinguish from one another. However, based on estimations from the pristine sample shown in **Figure 1(d & e)**, the analyzed area likely encompasses several tens of particles. In total, four SPED, seven EELS, and four EDXS datasets were acquired at different locations, ensuring that the observed features and conclusions drawn are statistically meaningful. A total of three TEM samples were analyzed from the first delithiated sample, derived from two different silicon electrodes, ensuring sufficient statistical relevance. Additionally, three attempts were made to prepare TEM samples after the 10$^{th}$ lithiation using two different electrodes. However, none were successful due to severe cracking, which obstructed effective lamella preparation. The cracking in the samples after the 10$^{th}$ lithiation primarily originates from the mud-type channel cracks formed during previous electrochemical cycles, which are further exacerbated by the peeling of the LPSCl layer. Two TEM samples from silicon after the 10$^{th}$ delithiation, obtained from a single electrode, were successfully analyzed. Sufficient regions were examined to ensure statistical relevance.

Despite the moderate statistical scope, the observed features remain consistent across available samples, and their transformations through different cycles are relatable. Given that such a study was previously impossible despite extensive research on silicon anodes (due to a lack of appropriate equipment) and is still difficult due to the tremendous amount of resources required, we regard these findings as significant. They provide a foundation for further studies and encourage reproducibility by other researchers in this field.

## Acknowledgments

Funding from the European Regional Development Fund (ERDF) and the Recovery Assistance for Cohesion and the Territories of Europe (REACT-EU), as well as the Bundesministerium für Bildung und Forschung (BMBF) within the FESTBATT cluster of competence (project 03XP0433D) and project SILKOMPAS (project 03XP0486D) is gratefully acknowledged. We also acknowledge the support of the German Research Foundation (DFG) via INST 160/724-1FUGG.

## Declaration of AI Assistance in Writing

The authors used ChatGPT for paraphrasing the original text, grammar correction, language refinement, and coherence improvement. They thoroughly reviewed and revised the content to ensure accuracy and originality, taking full responsibility for the final publication.

# References

[1]   J. Janek, W. G. Zeier, *Nat. Energy* **2016**, *1*, 1.
[2]   J. Sakabe, N. Ohta, T. Ohnishi, K. Mitsuishi, K. Takada, *Commun. Chem.* **2018**, *1*, 1.
[3]   M. Häusler, O. Stamati, C. Gammer, F. Moitzi, R. J. Sinojiya, J. Villanova, B. Sartory, D. Scheiber, J. Keckes, B. Fuchsbichler, S. Koller, R. Brunner, *Commun. Mater.* **2024**, *5*, 1.
[4]   K. Feng, M. Li, W. Liu, A. G. Kashkooli, X. Xiao, M. Cai, Z. Chen, *Small* **2018**, *14*, 1702737.
[5]   U. Kasavajjula, C. Wang, A. J. Appleby, *J. Power Sources* **2007**, *163*, 1003.
[6]   Z.-L. Xu, X. Liu, Y. Luo, L. Zhou, J.-K. Kim, *Prog. Mater. Sci.* **2017**, *90*, 1.
[7]   J. Sung, J. Heo, D.-H. Kim, S. Jo, Y.-C. Ha, D. Kim, S. Ahn, J.-W. Park, *Mater. Chem. Front.* **2024**, *8*, 1861.
[8]   S. Jayasubramaniyan, S. Kim, M. Ko, J. Sung, *ChemElectroChem* **2024**, *11*, e202400219.
[9]   X. Zhan, M. Li, S. Li, X. Pang, F. Mao, H. Wang, Z. Sun, X. Han, B. Jiang, Y.-B. He, M. Li, Q. Zhang, L. Zhang, *Energy Storage Mater.* **2023**, *61*, 102875.
[10]  B. Liang, Y. Liu, Y. Xu, *J. Power Sources* **2014**, *267*, 469.
[11]  D. Ma, Z. Cao, A. Hu, *Nano-Micro Lett.* **2014**, *6*, 347.
[12]  H. Wu, G. Chan, J. W. Choi, I. Ryu, Y. Yao, M. T. McDowell, S. W. Lee, A. Jackson, Y. Yang, L. Hu, Y. Cui, *Nat. Nanotechnol.* **2012**, *7*, 310.
[13]  Y. He, L. Jiang, T. Chen, Y. Xu, H. Jia, R. Yi, D. Xue, M. Song, A. Genc, C. Bouchet-Marquis, L. Pullan, T. Tessner, J. Yoo, X. Li, J.-G. Zhang, S. Zhang, C. Wang, *Nat. Nanotechnol.* **2021**, *16*, 1113.
[14]  E. Feyzi, A. K M r, X. Li, S. Deng, J. Nanda, K. Zaghib, *Energy* **2024**, *5*, 100176.
[15]  H. Nagata, K. Kataoka, *J. Power Sources* **2024**, *623*, 235443.
[16]  H. Nagata, J. Akimoto, K. Kataoka, *New J. Chem.* **2023**, *47*, 8479.
[17]  H. Huo, M. Jiang, Y. Bai, S. Ahmed, K. Volz, H. Hartmann, A. Henss, C. V. Singh, D. Raabe, J. Janek, *Nat. Mater.* **2024**, 1.
[18]  D. H. S. Tan, Y.-T. Chen, H. Yang, W. Bao, B. Sreenarayanan, J.-M. Doux, W. Li, B. Lu, S.-Y. Ham, B. Sayahpour, J. Scharf, E. A. Wu, G. Deysher, H. E. Han, H. J. Hah, H. Jeong, J. B. Lee, Z. Chen, Y. S. Meng, *Science* **2021**, *373*, 1494.
[19]  Z. Fan, B. Ding, Z. Li, Z. Chang, B. Hu, C. Xu, X. Zhang, H. Dou, X. Zhang, *eTransportation* **2023**, *18*, 100277.
[20]  M. J. Chon, V. A. Sethuraman, A. McCormick, V. Srinivasan, P. R. Guduru, *Phys. Rev. Lett.* **2011**, *107*, 045503.
[21]  S. Zhu, B. Lu, B. Rui, Y. Song, J. Zhang, *Electrochemistry* **2023**, *91*, 127002.
[22]  M. Otoyama, N. Terasaki, T. Takeuchi, T. Okumura, K. Kuratani, *ChemElectroChem* **2025**, *12*, e202400616.
[23]  Z. Fan, B. Ding, Z. Li, Z. Chang, B. Hu, C. Xu, X. Zhang, H. Dou, X. Zhang, *eTransportation* **2023**, *18*, 100277.
[24]  D. H. S. Tan, Y.-T. Chen, H. Yang, W. Bao, B. Sreenarayanan, J.-M. Doux, W. Li, B. Lu, S.-Y. Ham, B. Sayahpour, J. Scharf, E. A. Wu, G. Deysher, H. E. Han, H. J. Hah, H. Jeong, J. B. Lee, Z. Chen, Y. S. Meng, *Science* **2021**, *373*, 1494.
[25]  D. L. Nelson, S. E. Sandoval, J. Pyo, D. Bistri, T. A. Thomas, K. A. Cavallaro, J. A. Lewis, A. S. Iyer, P. Shevchenko, C. V. Di Leo, M. T. McDowell, *ACS Energy Lett.* **2024**, *9*, 6085.

[26] D. Qian, C. Ma, K. L. More, Y. S. Meng, M. Chi, *NPG Asia Mater.* **2015**, *7*, e193.
[27] S. Ahmed, A. Pokle, S. Schweidler, A. Beyer, M. Bianchini, F. Walther, A. Mazilkin, P. Hartmann, T. Brezesinski, J. Janek, K. Volz, *ACS Nano* **2019**, *13*, 10694.
[28] S. Ahmed, M. Bianchini, A. Pokle, M. S. Munde, P. Hartmann, T. Brezesinski, A. Beyer, J. Janek, K. Volz, *Adv. Energy Mater.* **2020**, *10*, 2001026.
[29] S. Ahmed, A. Pokle, M. Bianchini, S. Schweidler, A. Beyer, T. Brezesinski, J. Janek, K. Volz, *Matter* **2021**, *4*, 1.
[30] T. Demuth, T. Fuchs, F. Walther, A. Pokle, S. Ahmed, M. Malaki, A. Beyer, J. Janek, K. Volz, *Matter* **2023**, *6*, 2324.
[31] M. Malaki, J. Haust, J. P. Beaupain, H. Auer, A. Beyer, K. Wätzig, M. Kusnezoff, K. Volz, *Adv. Mater. Interfaces* **2023**, *10*, 2300513.
[32] M. Malaki, A. Pokle, S.-K. Otto, A. Henss, J. P. Beaupain, A. Beyer, J. Müller, B. Butz, K. Wätzig, M. Kusnezoff, J. Janek, K. Volz, *ACS Appl. Energy Mater.* **2022**, *5*, 4651.
[33] X. H. Liu, H. Zheng, L. Zhong, S. Huang, K. Karki, L. Q. Zhang, Y. Liu, A. Kushima, W. T. Liang, J. W. Wang, J.-H. Cho, E. Epstein, S. A. Dayeh, S. T. Picraux, T. Zhu, J. Li, J. P. Sullivan, J. Cumings, C. Wang, S. X. Mao, Z. Z. Ye, S. Zhang, J. Y. Huang, *Nano Lett.* **2011**, *11*, 3312.
[34] X. Wang, F. Fan, J. Wang, H. Wang, S. Tao, A. Yang, Y. Liu, H. Beng Chew, S. X. Mao, T. Zhu, S. Xia, *Nat. Commun.* **2015**, *6*, 8417.
[35] C. Shen, M. Ge, L. Luo, X. Fang, Y. Liu, A. Zhang, J. Rong, C. Wang, C. Zhou, *Sci. Rep.* **2016**, *6*, 31334.
[36] X. H. Liu, J. W. Wang, S. Huang, F. Fan, X. Huang, Y. Liu, S. Krylyuk, J. Yoo, S. A. Dayeh, A. V. Davydov, S. X. Mao, S. T. Picraux, S. Zhang, J. Li, T. Zhu, J. Y. Huang, *Nat. Nanotechnol.* **2012**, *7*, 749.
[37] C.-M. Wang, X. Li, Z. Wang, W. Xu, J. Liu, F. Gao, L. Kovarik, J.-G. Zhang, J. Howe, D. J. Burton, Z. Liu, X. Xiao, S. Thevuthasan, D. R. Baer, *Nano Lett.* **2012**, *12*, 1624.
[38] X. H. Liu, L. Zhong, S. Huang, S. X. Mao, T. Zhu, J. Y. Huang, *ACS Nano* **2012**, *6*, 1522.
[39] M. T. McDowell, I. Ryu, S. W. Lee, C. Wang, W. D. Nix, Y. Cui, *Adv. Mater.* **2012**, *24*, 6034.
[40] M. T. McDowell, S. Woo Lee, C. Wang, Y. Cui, *Nano Energy* **2012**, *1*, 401.
[41] M. Boniface, L. Quazuguel, J. Danet, D. Guyomard, P. Moreau, P. Bayle-Guillemaud, *Nano Lett.* **2016**, *16*, 7381.
[42] M. Ezzedine, F. Jardali, I. Florea, M.-R. Zamfir, C.-S. Cojocaru, *Batter. Supercaps* **2023**, *6*, e202200451.
[43] H. Valencia, P. Rapp, M. Graf, J. Mayer, H. A. Gasteiger, *J. Electrochem. Soc.* **2024**, *171*, 120507.
[44] J. Danet, T. Brousse, K. Rasim, D. Guyomard, P. Moreau, *Phys. Chem. Chem. Phys.* **2009**, *12*, 220.
[45] J.-M. Costantini, J. Ribis, *Materials* **2023**, *16*, 7610.
[46] H. Koh, E. Detsi, E. A. Stach, *Microsc. Microanal.* **2023**, *29*, 1350.
[47] E. V. Astrova, A. M. Rumyantsev, G. V. Li, A. V. Nashchekin, D. Yu. Kazantsev, B. Ya. Ber, V. V. Zhdanov, *Semiconductors* **2016**, *50*, 963.
[48] E. Sivonxay, M. Aykol, K. A. Persson, *Electrochimica Acta* **2020**, *331*, 135344.
[49] G. Lener, M. Otero, D. E. Barraco, E. P. M. Leiva, *Electrochimica Acta* **2018**, *259*, 1053.
[50] S. Pidaparthy, H. Ni, H. Hou, D. P. Abraham, J.-M. Zuo, *Ultramicroscopy* **2023**, *248*, 113718.

[51] A. V. Martin, E. D. Bøjesen, T. C. Petersen, C. Hu, M. J. Biggs, M. Weyland, A. C. Y. Liu, *Small* **2020**, *16*, 2000828.

[52] R. K. Biswas, J. Ghosh, M. Kuttanellore, *Trans. Indian Ceram. Soc.* **2020**, *79*, 158.

[53] F. Zhu, J. Singh, *J. Appl. Phys.* **1993**, *73*, 4709.

[54] L. Ley, in *Phys. Hydrog. Amorph. Silicon II Electron. Vib. Prop.* (Eds.: J. D. Joannopoulos, G. Lucovsky), Springer, Berlin, Heidelberg, **1984**, pp. 61–168.

[55] S. Lin, X. Li, J. Zhang, L. Guo, X. Xu, W. Yuan, H. Liu, A. Li, X. Chen, H. Song, *J. Colloid Interface Sci.* **2025**, *699*, 138302.

[56] Y. Huang, S. Jing, H. Shen, S. Li, Y. Shen, Y. Lin, Y. Zhang, Z. Zhang, Y. Liu, Y. Chen, F. Liu, Y. Lu, *Acta Mater.* **2025**, *289*, 120915.

[57] S.-Y. Ham, E. Sebti, A. Cronk, T. Pennebaker, G. Deysher, Y.-T. Chen, J. A. S. Oh, J. B. Lee, M. S. Song, P. Ridley, D. H. S. Tan, R. J. Clément, J. Jang, Y. S. Meng, *Nat. Commun.* **2024**, *15*, 2991.

[58] S. Roychoudhury, Z. Zhuo, R. Qiao, L. Wan, Y. Liang, F. Pan, Y. Chuang, D. Prendergast, W. Yang, *ACS Appl. Mater. Interfaces* **2021**, *13*, 45488.

[59] F. de la Peña, E. Prestat, J. Lähnemann, V. T. Fauske, P. Burdet, P. Jokubauskas, T. Furnival, C. Francis, M. Nord, T. Ostasevicius, K. E. MacArthur, D. N. Johnstone, M. Sarahan, J. Taillon, T. Aarholt, pquinn-dls, V. Migunov, A. Eljarrat, J. Caron, T. Nemoto, T. Poon, S. Mazzucco, actions-user, sivborg, N. Tappy, N. Cautaerts, S. Somnath, T. Slater, M. Walls, pietsjoh, **2024**, DOI 10.5281/zenodo.14057415.

# Supplementary Information

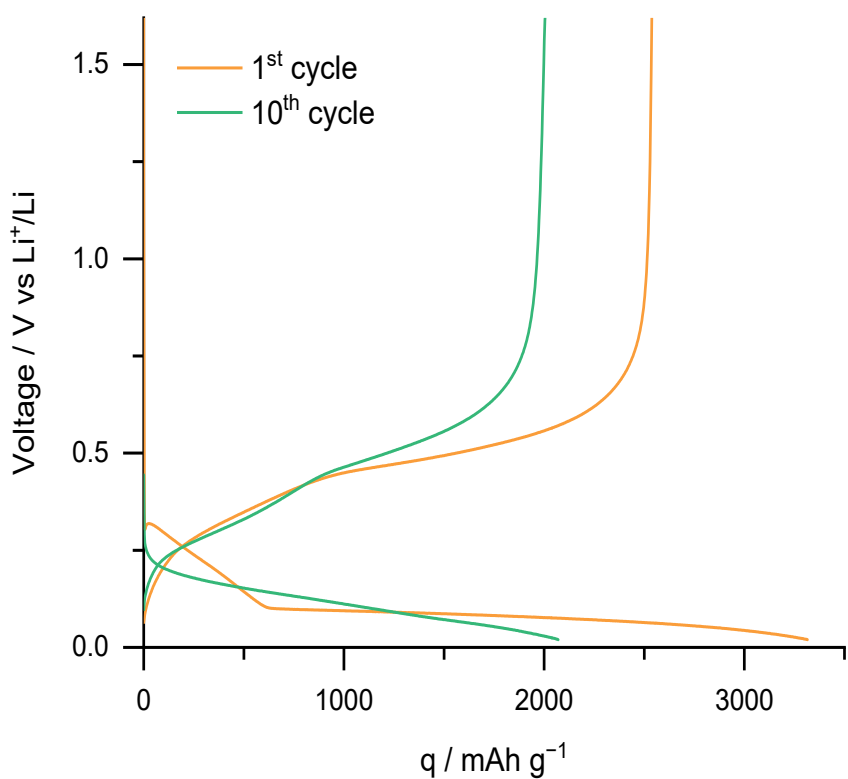

**Figure S1:** First/tenth charging curve alongside the first and tenth discharge curves. The silicon sheet was cycled in a pressed-cell configuration using In(InLi)$_x$ as the counter electrode, with a stack pressure of ≈60 MPa, and a fixed current of 0.48 mA, which correspond to a C-rate of C/10. An open circuit voltage (OCV) step of 5 hours was performed before cycling to let the InLi alloy form and a 5-minute OCV step between every galvanostatic charge and discharge was added.

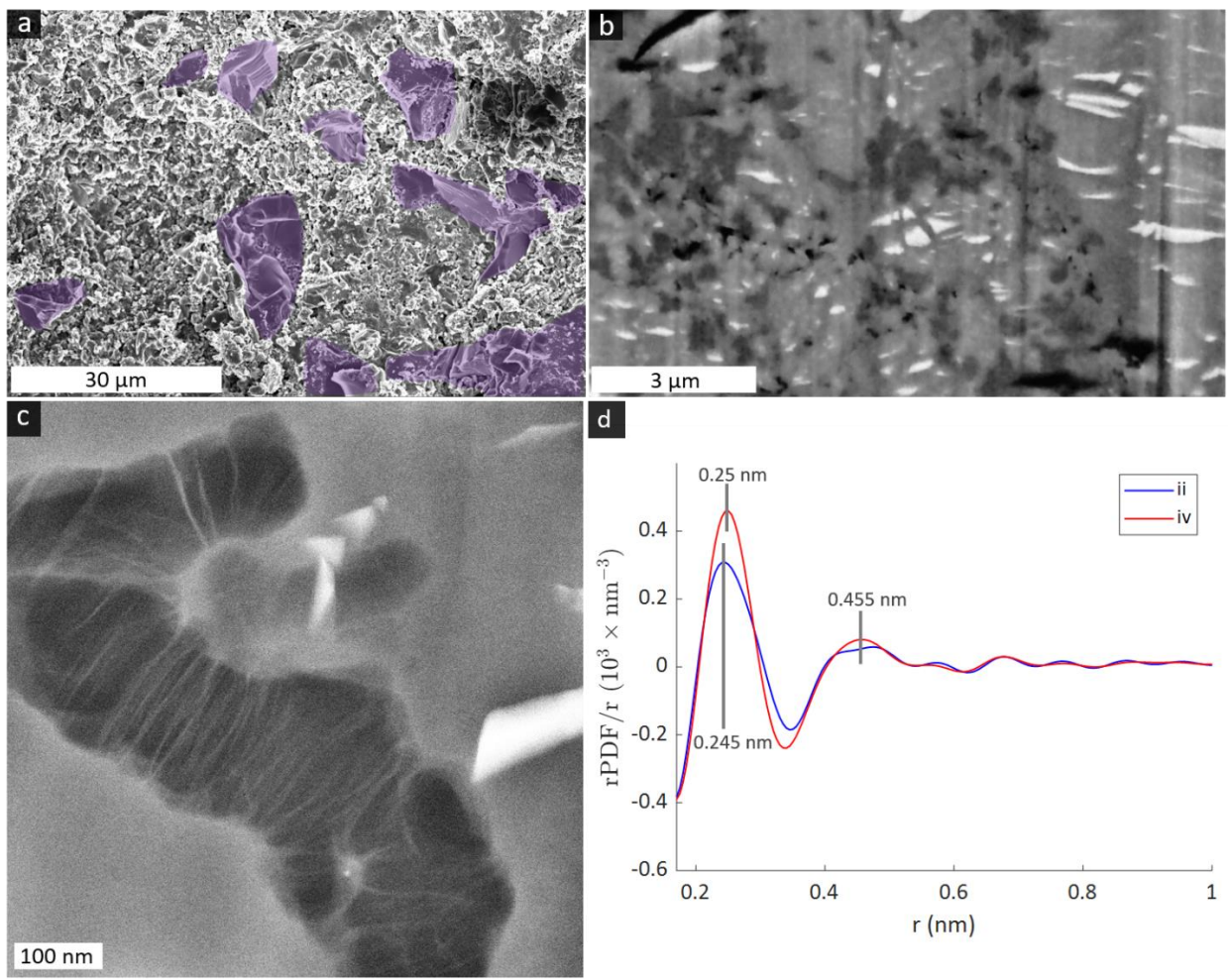

**Figure S2:** (a) Surface of the silicon electrode after the first lithiation and removing bulk $Li_6PS_5Cl$ layer, showing residual traces of $Li_6PS_5Cl$ exemplarily marked by purple color. (b) Cryo-secondary electron image of the cross-section during TEM sample preparation of the silicon electrode after first lithiation. (c) a Cryo-STEM HAADF image of an area in the sample after first lithiation showing dendrite-like features. (d) Radial pair distribution function (rPDF) derived from the diffraction patterns 'ii' and 'iv', shown in **Figure 2**.

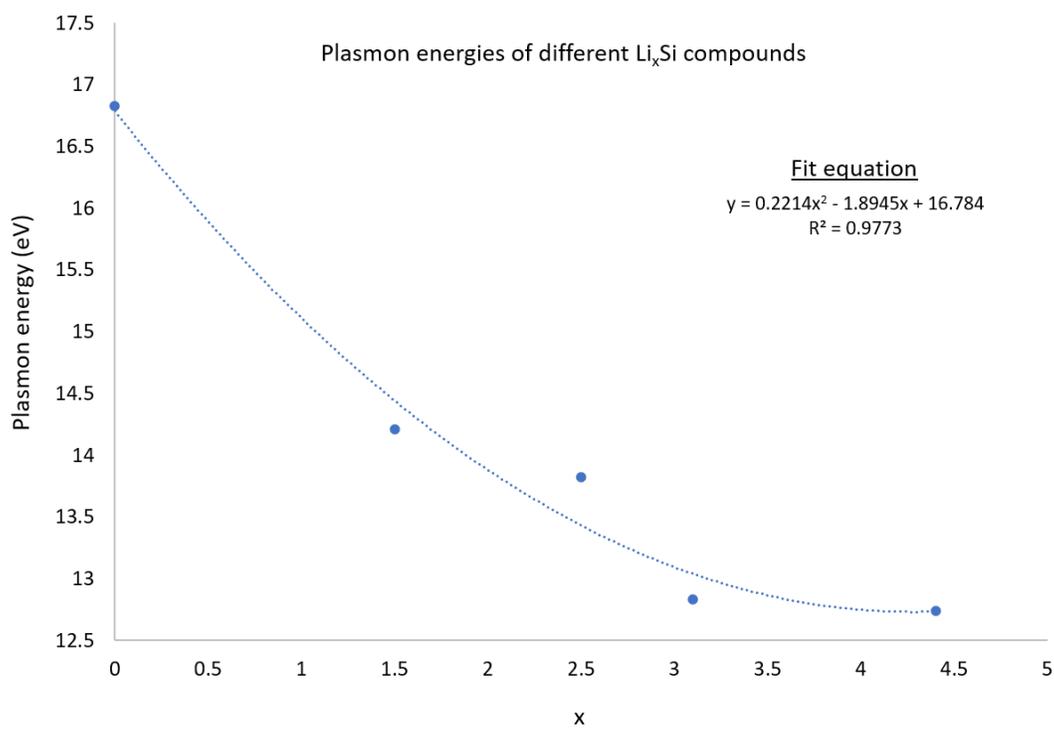

**Figure S3:** Fitting curve and associated equation for plasmon energies of Li$_x$Si alloys (x = 0 to x = 4.4), used to analyze the *n*(Li):*n*(Si) ratios in the silicon electrode.

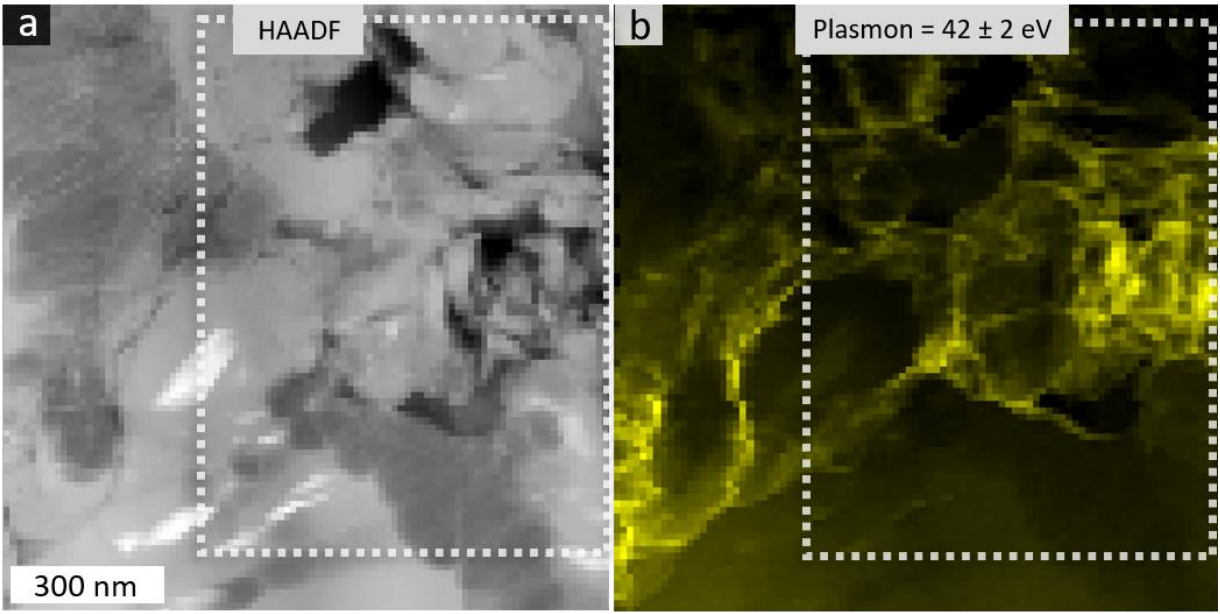

**Figure S4:** (a) Cryo-STEM HAADF image highlighting the microstructure of the silicon electrode after 1$^{st}$ charge. (b) Cryo valence electron EELS map constructed using the 42 eV ± 2 eV integration window after Lucy-Richardson deconvolution, showing significant concentrations of $Li_2O$ and lithium-silicate compounds at the grain boundaries. Bright regions qualitatively show relatively higher amounts of the $Li_2O$ and lithium-silicate compounds.

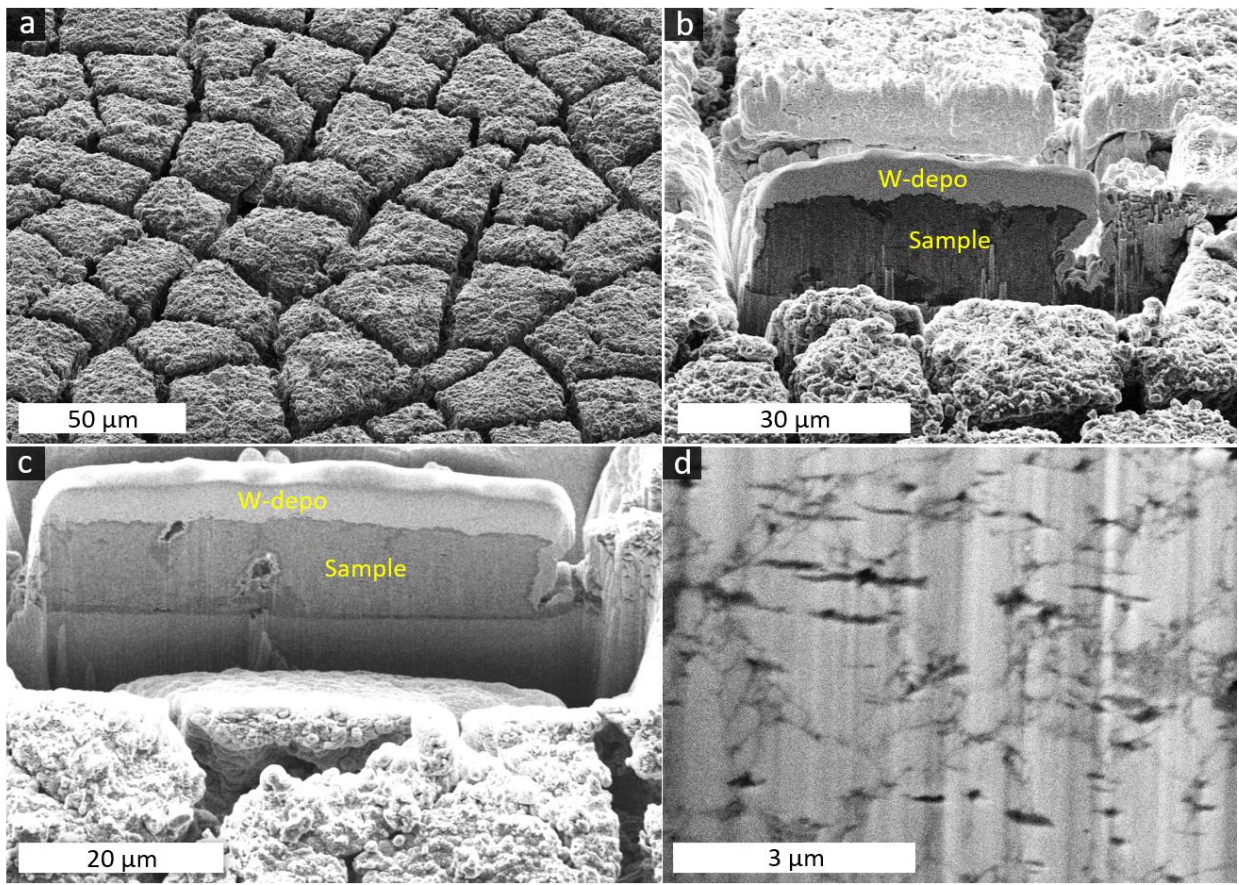

**Figure S5:** (a) Top surface of the silicon electrode after the first discharge, showing mud-type channel cracks. (b, c) Cross-sectional views of the silicon electrode during the lift-out process. (d) Magnified cross-sectional image highlighting partial porosity recovery.

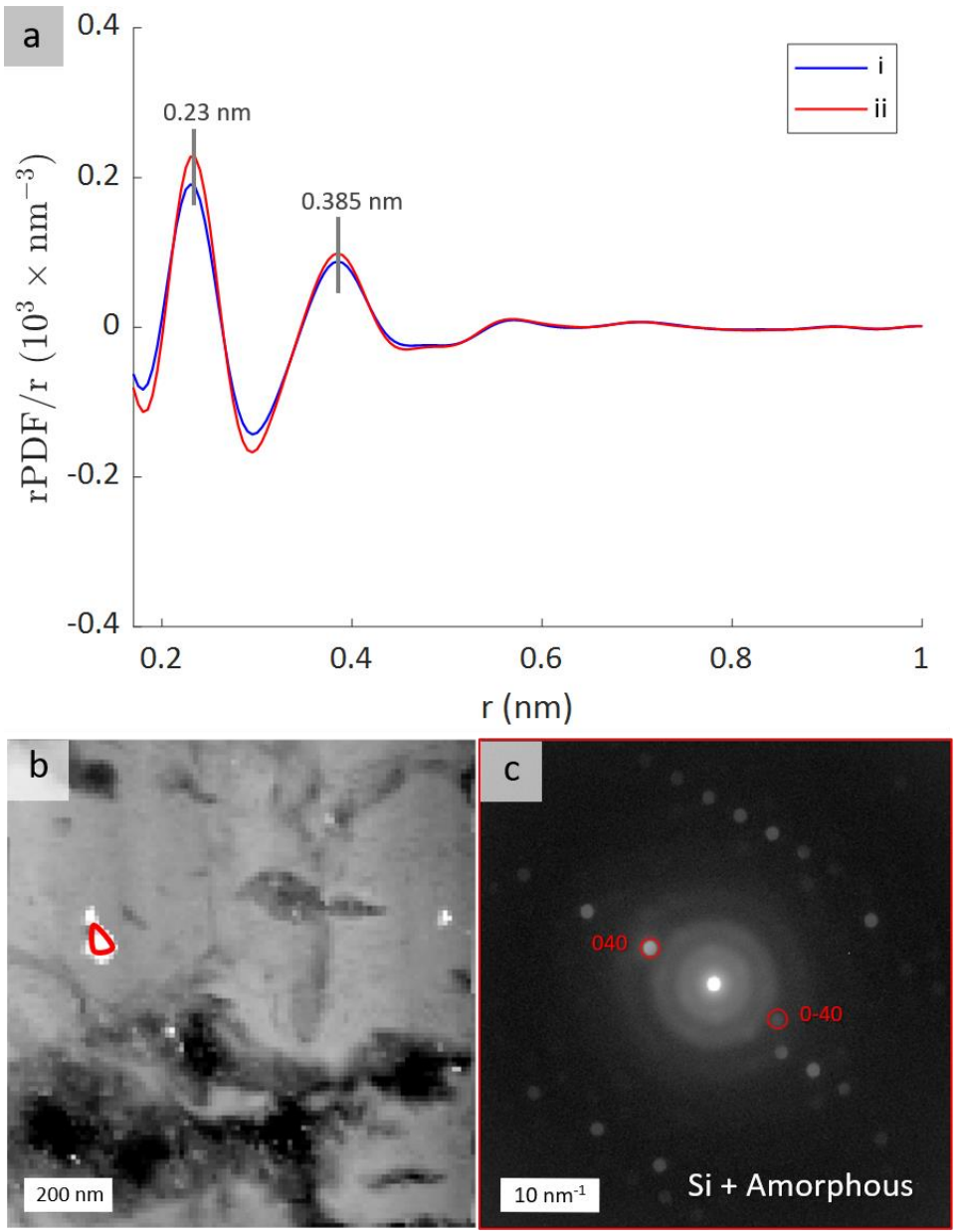

**Figure S6:** (a) Radial pair distribution functions (rPDFs) for denser 'i' and less dense 'ii' regions (shown in Figure 4(c)) of the amorphous silicon region. (b) Virtual dark-field image from a 4D SPED dataset in a region after 1st discharge. (c) Position-averaged SNBD pattern from a region marked by a red line in (b). Some of the reflections respective to the silicon crystalline structure are exemplarily marked in (c).

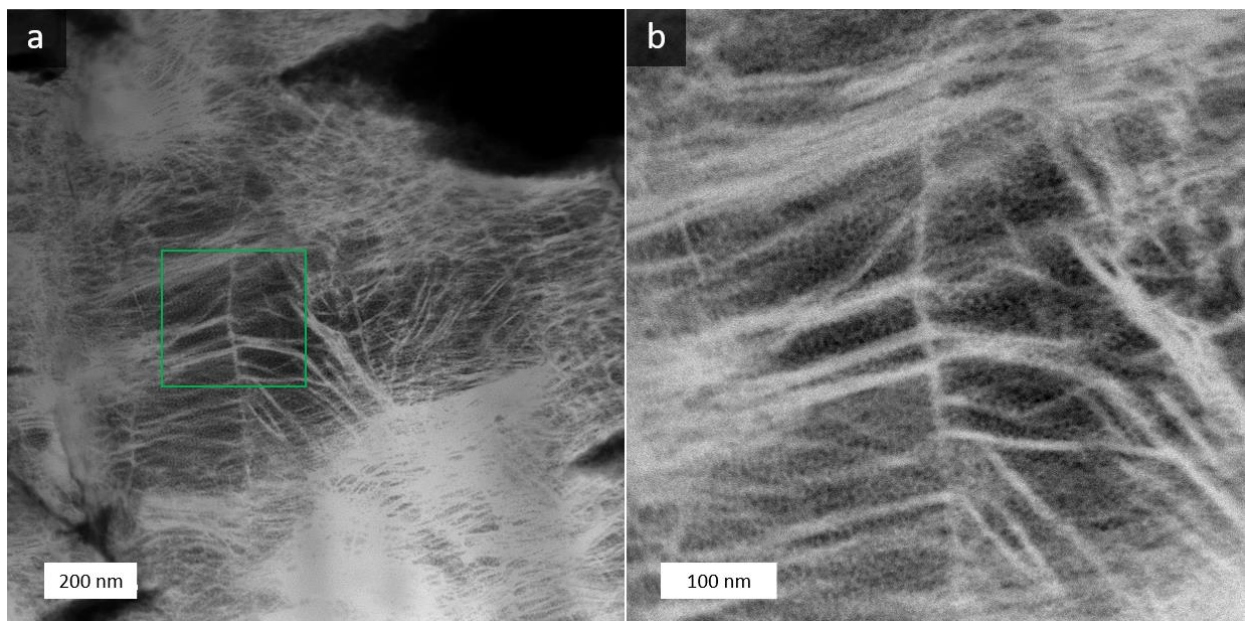

**Figure S7:** (a) Cryo-STEM HAADF images of a thread-like, irregularly meshed region. (b) Magnified cryo-STEM HAADF image of the area highlighted in (a), which was also utilized for cryo-STEM HAADF tomography shown in **Videos SV2 and SV3**.

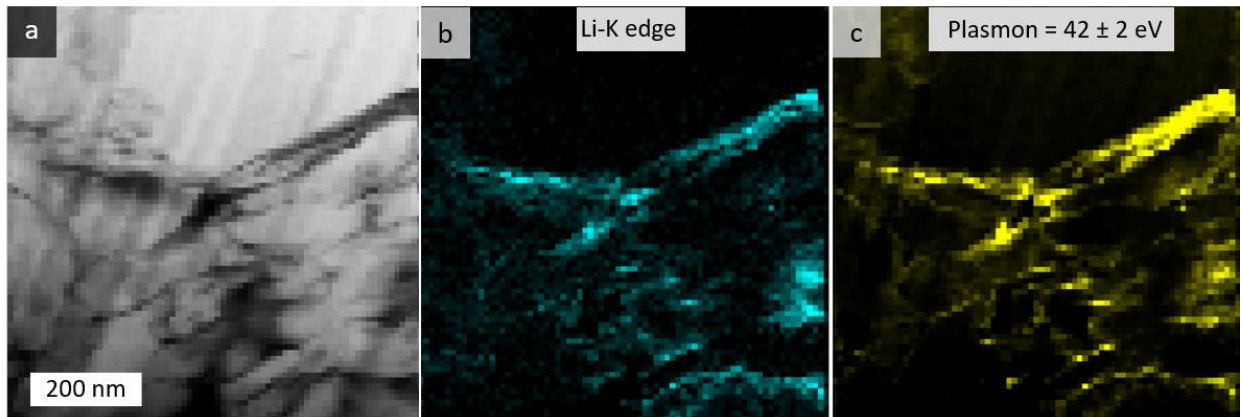

**Figure S8:** (a) Cryo-STEM HAADF image of a region after the first discharge. (b) Cryo-STEM EELS map, constructed by integrating the energy window around the Li peak (60 eV ± 2 eV). (c) Cryo-STEM valence EELS map using the second plasmon of the $Li_2O$-$SiO_2$ mixture (42 eV ± 2 eV). Bright regions qualitatively show relatively higher local amounts of the $Li_2O$ and lithium-silicate compounds.

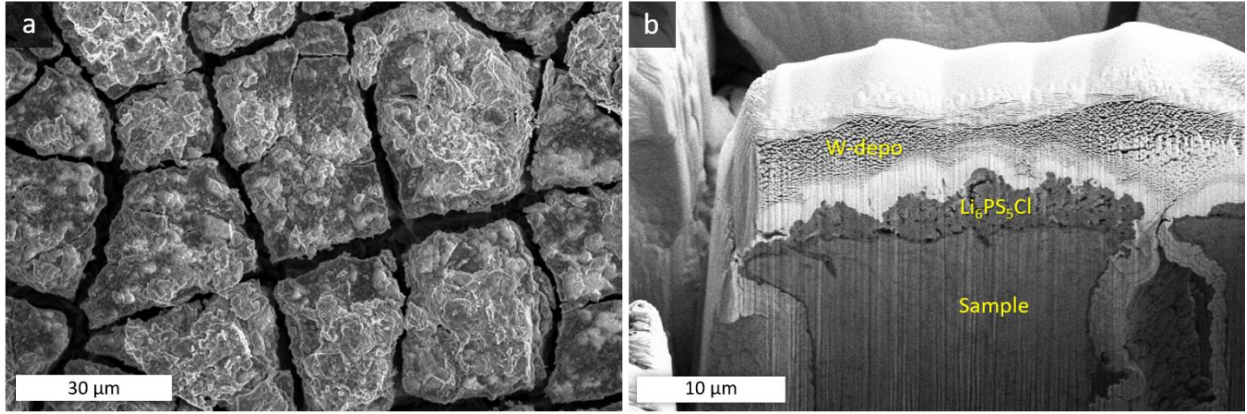

**Figure S9:** (a) SEM image of the silicon electrode surface after 10 cycles, showing the presence of mud-type channel cracks. (b) SEM image of the cross-section during the lift-out process.

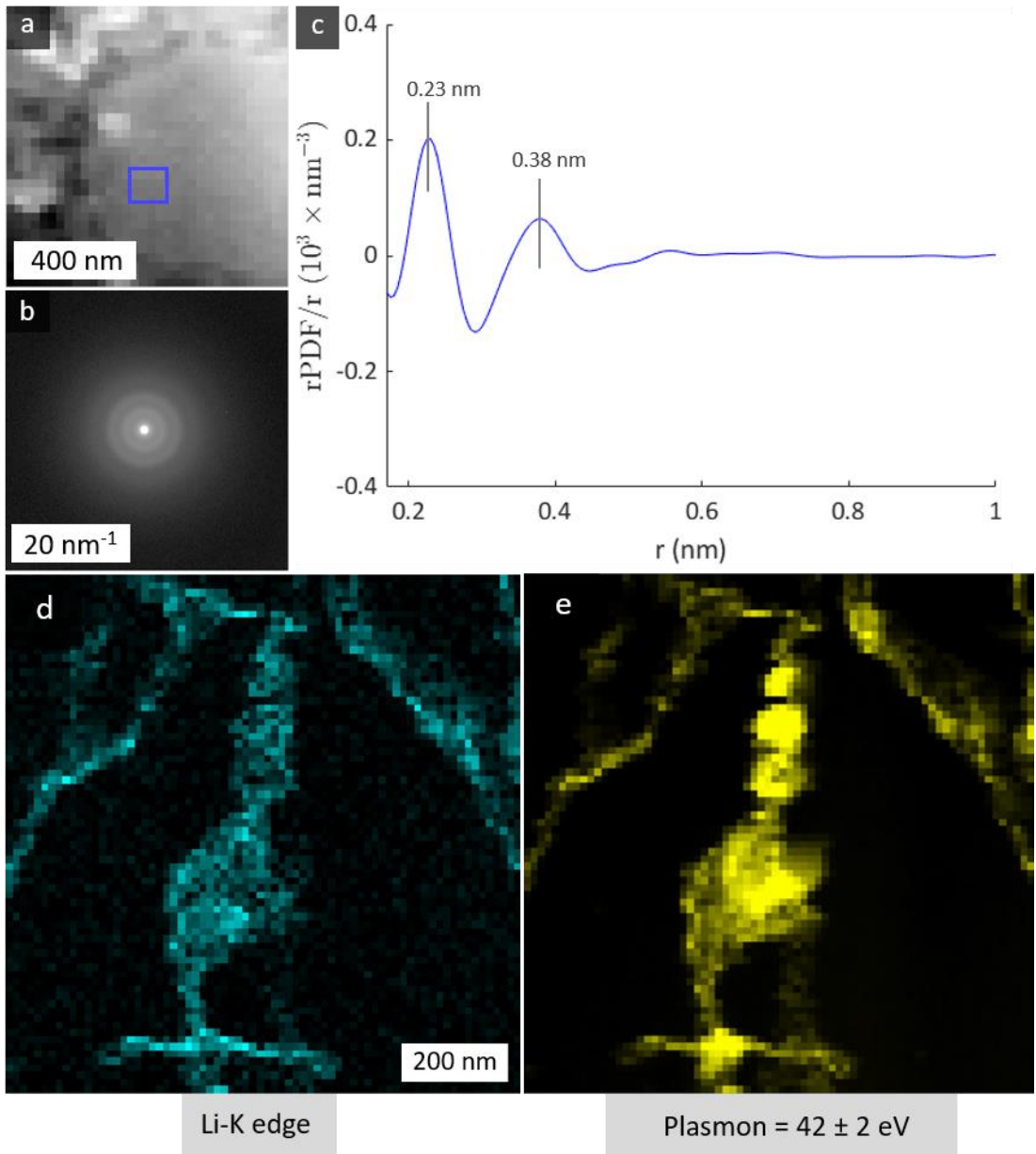

**Figure S10:** (a) Virtual dark-field image derived from a 4D SPED dataset. (b) Position-averaged SPED pattern from the region marked in (a). (c) rPDF corresponds to the pattern in (b). (d, e) Cryo-STEM EELS maps were generated using the lithium-K edge and the second plasmon of the $Li_2O$-$SiO_2$ mixture (42 eV ± 2 eV), respectively. Bright regions qualitatively show relatively higher amounts of the $Li_2O$ and lithium-silicate compounds.